\documentclass[twocolumn,showpacs,preprintnumbers]{revtex4}
\usepackage{dcolumn}
\usepackage{bm}
\input epsf

\begin{document}

\preprint{astro-ph/0812.3865}

\title{Accelerated expansion and matter creation}

\author{V\'{\i}ctor H. C\'{a}rdenas}
\email{victor@dfa.uv.cl}

\affiliation{Departamento de F\'{\i}sica y Astronom\'{i}a,
Universidad de Valpara\'{i}so, Av. Gran Bretana 1111,
Valpara\'{i}so, Chile}

\begin{abstract}
A set of cosmological models that takes into account the variation
of the particle number is presented. In this context both dark
matter and dark energy can be explained using a single component,
without assuming any exotic equation of state, solving directly the
cosmic coincidence problem.
\end{abstract}


\maketitle

\section{Introduction}

Currently the observational evidence coming from supernovae studies
\cite{SNIa}, cosmic background radiation fluctuations \cite{cmbr},
and baryon acoustic oscillations \cite{bao}, set a strong case for a
cosmological (concordance) universe model composed by nearly $70$
percent of a mysterious component called dark energy, responsible
for the current accelerated expansion, nearly $25$ percent of dark
matter, which populate the galaxies halos and a small percent
(around $4$) composed by baryonic matter. The nature of these two
dark component remains so far obscure \cite{DEDM}. In the case of
dark matter, we have a set of candidates to be probed with
observations and detections in particle accelerators, however the
case of dark energy is more elusive. This component not only has to
fill the universe at the largest scale homogeneously, and to have
the appropriate order of magnitude to be comparable with that of
dark matter, but also has to have a negative pressure equation of
state, something that was assumed first to explain inflation in the
early universe, inspired by high energy theories of particle
physics, but which seems to be awkward to appeal for, at these low
energy scales. The alternative way used to account for the cosmic
acceleration is to consider a modification of general relativity at
large scales \cite{alterGR}. However it seems to be a difficult
route to follow, considering the subtleties in the interpretation of
the observations even in the simplest model based on general
relativity.

To describe the universe at the largest scale, as is the case in
cosmology, we have to make certain assumptions regarding the local
distribution of matter and its characteristics. As is well known
(and explained in several classic books \cite{tolman},
\cite{weinberg}, \cite{peebles}) the energy-momentum tensor used in
the Einstein's equations is imported from considerations made in
special relativity, through a covariant generalization supported by
the equivalence principle. The assumption about homogeneity and
isotropy leads to consider the matter content of the universe as a
fluid. In the case of a perfect fluid, defined as a mechanical
medium incapable of exerting transverse stresses, it takes the
standard diagonal form
\begin{equation}\label{temunu}
T^{\mu \nu} = (\rho + p)U^{\mu}U^{\nu}-g^{\mu \nu}p,
\end{equation}
where $\rho$ is identified as the energy density of the fluid, $p$
is the proper hydrostatic pressure, $U^{\mu}=dx^{\mu}/ds$ are the
components of the macroscopic velocity of the fluid with respect to
the actual coordinate system and $g^{\mu \nu}$ is the metric of it.
The expression (\ref{temunu}) is what is measured by an observer at
rest in the fluid, who examines an element of the fluid small enough
so that gravitational curvature can be neglected. The conclusion of
applying the energy conservation relation to this tensor
\begin{equation} \label{conslaw}
T^{\mu\nu}_{\hspace{0.3cm};\nu}=0,
\end{equation}
is that perfect fluids behave adiabatically when examined by a local
observer, satisfying the equation
\begin{equation}\label{ds0}
d(\rho V)+ pdV=0,
\end{equation}
where $V$ is the physical volume of the region considered. Is usual
to take as the volume $V =d_f^{3}$, where $d_f$ is a physical
distance of the order of the region we are interested in. Using that
$d_f = d_c R(t)$, leads to the well know result
\begin{equation}
\dot{\rho}+3\frac{\dot{R}}{R}(\rho + p)=0,
\end{equation}
To get this result, it was assumed also that particle number is
conserved $N^{\mu}_{;\mu}=0$, where $N^{\mu}=nU^{\mu}$ with $n$ the
particle number density.

A way to modify (\ref{ds0}) without spoiling adiabatic evolution is
by adding a work term due to the change of the particle number $N$.
In \cite{prigogine}, the authors studied this case as an alternative
cosmological model, considering the change of the particle number,
assuming that such a correction have to be considered during matter
creation. This work actually suggested for the first time, a way to
incorporate the particle creation process in the context of
cosmology, in a self consistent way. In fact, the original claim
made by Zeldovich \cite{zeldov70}, that gravitational particle
production can be described phenomenologically by a negative
pressure, is realize here in a beautiful way. A covariant
formulation of the model was presented in \cite{CLW90}.

The context where these considerations would be important were
mentioned to be; the steady-state cosmological model, the warm
inflationary scenario and within the standard inflationary scenario,
during the reheating phase. Based on this work, the studies of
cosmological models with matter creation \cite{lima96} were
initiated, which were rapidly recognized as be potentially important
to explain dark energy \cite{harko}. In particular, in
Ref.\cite{zsbp01} the authors established a model where dark energy
can be mimicked by self-interactions of the dark matter substratum.
Within the same framework, models of interacting dark energy and
dark matter were proposed \cite{interact}. Actually, in these
studies is possible to have consistently, a universe where matter
creation proceeds within an adiabatic evolution. More recently, Lima
et al.,\cite{lima08} have presented a study of a flat cosmological
model where a transition from decelerated to accelerated phase
exist. They explicitly show that previous models considered
\cite{lima99}, does not exhibit the transition, and study the
observational constraints on the model parameters.

In this work I consider non flat models where by modifying the first
law, enable us to explain the current acceleration of the universe
expansion through a fictitious pressure component coming from
changes in the total particle number. In this way, the dark energy
component is basically an effect due to the dark matter creation
process in the universe, fact that enable us to explain easily the
cosmic coincidence problem; it is not strange to have a similar
contribution from these two component, because they are just one
component; dark matter. I also demonstrate that a non constant
matter creation rate can indeed lead to a transition from a
decelerated to an accelerated regime even in the case of non flat
geometries. In the next section I derive the equations of motion in
the case of matter creation. Then, I open section III with a very
simple model with a non constant creation rate, that resemble the
$\Lambda$CDM model. After this, I discussed the case of the
transition from a decelerated to an accelerated expansion in non
flat models.

\section{Matter creation}

The incorporation of the variable $N$ into the cosmological
scenario, modifies the first law as
\begin{equation}\label{firstlaw}
d(\rho V)+ pdV -(h/n)d(nV)=0,
\end{equation}
where $h$ is the enthalpy (per unit volume), $n=N/V$ is the number
density and $\rho$ is the energy density. The important thing to
stress here is that the universe evolution continues to be
adiabatic, in the sense that the entropy per particle remains
unchanged ($\dot{\sigma}=0$). The extra contribution can be
interpreted as a non thermal pressure defined as
\begin{equation}\label{pc}
p_c=-\frac{h}{n}\frac{d(nV)}{dV}.
\end{equation}
This is the source that produces the current acceleration of the
universe expansion. Once the particle number increases with the
volume ($dN/dV>0$), we obtain a negative pressure.

Now, because we are considering matter creation, we have to impose
the second law constraint as
\begin{equation}\label{second}
dS=\frac{s}{n}d(nV) \geq 0,
\end{equation}
where $s=S/V$ is the entropy density. From (\ref{firstlaw}) we find
that the new set of Einstein equations are
\begin{equation}\label{friedman}
  H^2 + \frac{k}{R^2} =  \frac{8\pi G}{3} \rho
\end{equation}
\begin{equation}\label{rho}
  \dot{\rho} = \frac{\dot{n}}{n}(\rho + p)
\end{equation}
and from (\ref{second}) we have to supply an extra relation between
$n$ and the Hubble parameter
\begin{equation}\label{ene}
  \dot{n} + 3 H n = 3H\beta,
\end{equation}
where $\beta$ is a definite positive function. The set of equations
(\ref{friedman}), (\ref{rho}) and (\ref{ene}) completely specified
the system evolution. The standard adiabatic evolution is easily
recovered: setting $\beta = 0$ implies that $\dot{n}/n=-3H$, which
leads to the usual conservation equation from (\ref{rho}). It is
interesting to note that Eq.(\ref{rho}) enable us to determine the
pressure; for example if $\rho=m n$ Eq.(\ref{rho}) implies $p=0$,
and furthermore if $\rho=aT^4$ and $n=bT^3$ implies $p=\rho/3$.

\section{Models}

Let us assume first here that
\begin{equation}\label{asump1}
\frac{dN}{dV}= \beta_0
\end{equation}
The arguments in favor of this hypothesis would be given in short.
Let us study here its consequences. If we consider $V=R^3$, relation
(\ref{asump1}) leads to $\beta=\beta_0$. From (\ref{ene}) we finds
the solution for the density number to be
\begin{equation}\label{sol1}
n(t)= \beta_0 + \frac{C_1}{R(t)^3},
\end{equation}
where $C_1$ is a constant of integration. Assuming $\rho=m n$ as for
non-relativistic matter (that leads to que usual equation of state
$p=0$ as we discussed that at the end of the last section), we get
for the energy density
\begin{equation}\label{rhosol}
\rho(R)= m\beta_0 + \frac{mC_1}{R^3},
\end{equation}
that resembles the combined contribution of a cosmological constant
and dust. The level of fine tuning here to obtain an accelerated
expansion is relatively smaller than in the usual $\Lambda$CDM
model, because in this case, both terms come from the same function,
and we know that we can obtain an accelerated expansion phase after
some time from (\ref{friedman}). Note however that the equation of
state has not been modified, $\rho$ is still the non-relativistic
contribution. We do not have to introduce any exotic component -
with a negative pressure - to describe the current expansion
acceleration. Given the current status of the dark matter and dark
energy problem \cite{DEDM}, we can use this idea as a possible way
to understand it.

In terms of the exotic dark energy fluid with energy density
$\rho_X$ and pressure $p_X$, the solution (\ref{rhosol}) can be
expressed as
\begin{equation}\label{eq1}
\rho_X= m\beta, \hspace{1cm} p_X=\omega \rho_X,
\end{equation}
where now there must be two copies of Eq.(\ref{rho}) with the one
corresponding to dark energy being
\begin{equation}\label{de1}
 \dot{\rho_X} = \frac{\dot{n}}{n}(\rho_X + p_X),
\end{equation}
and $n$ stand only for the second term in Eq. (\ref{rhosol}). Of
course, from Eqs. (\ref{friedman}), (\ref{rhosol}) and (\ref{de1})
we get $\omega=-1$; e. i., the ansatz (\ref{asump1}) has leads us to
the case of a purely cosmological constant contribution. In general,
we can always separate the contribution for ``purely''
non-relativistic matter $\rho_m$ $(\propto R^{-3}$) from the dark
energy. In this case $\rho = \rho_m + \rho_X$, and the pressure can
be written as
\begin{equation}\label{px}
p_X= - m \beta_0.
\end{equation}
We can notice that $p_X$ is exactly the non thermal pressure
computed in (\ref{pc}) (recall that $p_m=0$).

I have to stress here that the main result derived in this section
means that we have a single contribution, which satisfy the
non-relativistic matter equation of state, that describe both dark
matter and dark energy simultaneously, and in this way solving
automatically the coincidence problem. This dark unification
mechanism does not have the problem studied in \cite{STZW}, because
in the cases studied in that paper, for example the Chaplygin gas
model \cite{chaplygin}, the sound velocity of the dark matter is not
zero, leading to instabilities. That happens because the Chaplygin
gas interpolates between the equation of states for dark matter and
dark energy. This does not happens here, because there is just one
equation of state; that of non-relativistic matter.

\section{Decelerated/accelerated transition}

As is well known from observations of supernovae, a transition from
a decelerated to an accelerated expansion occurs in the recent
history of the universe. Depending on what is used to model dark
energy, different redshift have been obtained between $0.5$ to $1$.

This topics was recently discussed in \cite{lima08} where the
authors specialized in a flat cosmology, where a explicit transition
can be achieved from a decelerated to an accelerated expansion. In
this section we generalize this work to non flat universes, by using
a model (see equation (\ref{ene})) with
\begin{equation}\label{rate1}
\beta = n\gamma \frac{H_0}{H}
\end{equation}
where $\gamma$ is a constant parameter (is exactly the same letter
introduced in \cite{lima08}). In what follows, a non relativistic
matter equation of state is assumed ($p=0$).

Introducing (\ref{rate1}) in (\ref{ene}) leads to the following
solution
\begin{equation}\label{nder}
n(R)=n_0\left(\frac{R_0}{R}\right)^3 e^{3H_0\gamma (t-t_0)}.
\end{equation}
Is evident the meaning of the subscript zero. Because $\rho=mn$ for
non-relativistic matter, the Friedman equation can be written as
\begin{equation}\label{req1}
\dot{R}^2R+kR = \kappa e^{3H_0\gamma (t-t_0)},
\end{equation}
where $\kappa=8\pi G \rho_0 R_0^3/3$. Evaluating this expression for
the reference time $t=t_0$ we obtain that
\begin{equation}\label{kdef}
    k=H_0^2 R_0^2 (\Omega_0 -1),
\end{equation}
where we have defined $\Omega_0=\rho/\rho_c$, with $\rho_c=3
H_0^2/8\pi G$ being the critical energy density. Using this in
(\ref{req1}) we obtain
\begin{equation}\label{fried2}
\frac{\dot{R}^2}{R_0^2}=H_0^2\left[ 1-\Omega_0 + \Omega_0 \left(
\frac{R_0}{R}\right) e^{3H_0\gamma (t-t_0)}\right].
\end{equation}
Clearly, for $\gamma=0$ this expression reduces to the standard
result.

In order to compare with observations, we shall compute the
deceleration parameter, $q=-R\ddot{R}/\dot{R}^2$, in terms of the
redshift. By differentiating (\ref{fried2}) respect to cosmic time
we find
\begin{equation}\label{qdet}
q = - \frac{(3H_0 \gamma - H)e^{3H_0\gamma (t-t_0)}}{2H\left[
1-\Omega_0 + \Omega_0 (\frac{R_0}{R})e^{3H_0\gamma (t-t_0)}\right]}.
\end{equation}
So, from (\ref{fried2}) we find the scale factor in term of the
cosmic time, $R(t)$, and using that $R=R_0 (1+z)^{-1}$, we can use
(\ref{qdet}) to write down $q(z)$. As an example, for $\Omega_0=1.2$
and $\gamma=0.4$ the deceleration parameter computed from
(\ref{qdet}) in terms of the redshift is shown in Figure 1.
\begin{figure}[h]
\centering \leavevmode\epsfysize=5cm \epsfbox{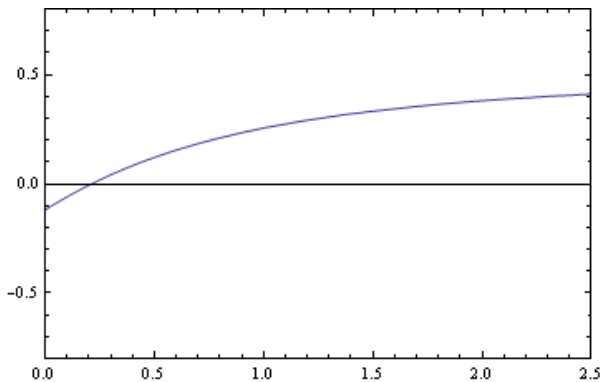}\\
\caption{\label{fig1} $q(z)$ versus $z$. An example of
Eq.(\ref{qdet}) in terms of the redshift $z$. In this case
$\Omega_0=1.2$ and $\gamma=0.4$. See the text for comments. }
\end{figure}
For a fixed $\Omega_0$, a crossing redshift exist if $\gamma >1/3$.
Otherwise, the universe never made the transition. For $\Omega_0<1$,
the situation is qualitatively similar, in this case with a smaller
slope.

In the case of a flat universe, $\Omega_0=1$ we obtain
\begin{equation}\label{qflat}
q_0 =  \frac{1}{2}(1-3\gamma \frac{H_0}{H}),
\end{equation}
which coincides with the result obtained in \cite{lima08}.

To complete the analysis, I will briefly discuss the case of the
interaction term $\Gamma=3\beta H$, discussed also in \cite{lima96}.
In this case, the deceleration parameter can be explicitly written
in terms of the redshift as
\begin{equation}\label{qz2}
q(z)=\left(\frac{1-3\beta}{2} \right)\left(\frac{\Omega_0
(1+z)^{1-3\beta}}{\Omega_0 (1+z)^{1-3\beta} +1-\Omega_0} \right).
\end{equation}
If we specialize to a flat universe, $\Omega_0=1$, $q$ becomes a
constant, showing no transition between decelerated to an
accelerated regime. However, the non flat case, is not different.
Actually, the $z$ dependence make the values of $q$ to vary with
redshift, but there is no redshift for which $q=0$.

So the challenge would be to estimate the function $dN/dV$ enabling
us to give the best fit to the observations. The first model
considered in the last section shows us a very simple scenery to
follow. Let me proceed further and complicate a little bit the model
setup. For example, we can study the case of a distribution of
matter which is oscillatory with volume. It means that instead of
consider (\ref{asump1}), we use $dN/dV=\beta \cos(V/V_c)$. This
implies that matter distribution is characterized by a volume $V_c$.
In this case, the energy density behaves as $\rho = C_1/R^{3}+(\beta
V_c/R^{3})\sin(R^{3}/V_c)$, so when the volume considered is small
enough $R^{3}\ll V_c$, this solution approaches (\ref{rhosol}). A
typical profile of the scale factor evolution is shown in Fig. 2.
\begin{figure}[tb]
\centering \leavevmode\epsfysize=5cm \epsfbox{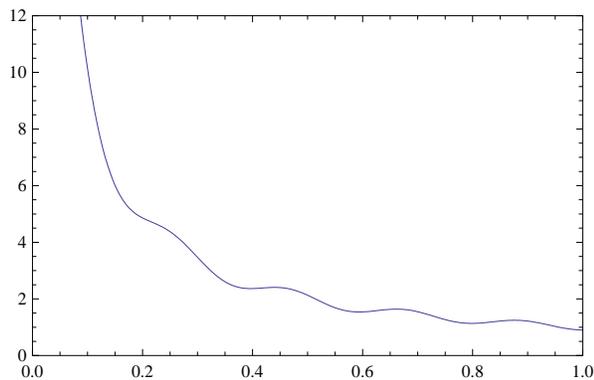}\\
\caption{\label{fig2} Typical profile of the scale factor evolution
in the case of a oscillatory distribution of matter. Although the
global behavior is $\propto R^{-3}$, the oscillatory nature of the
distribution of matter produces wiggles that can be associated to
short periods of accelerated expansions. The scales are arbitrary.}
\end{figure}


In this letter I have demonstrated that considering small changes in
the total number of particles in our universe, offer a possible way
to understand the current accelerated expansion measurements,
without using any exotic energy component. Also, this scenario
explain also the cosmic coincidence problem, basically showing that
these two components, dark matter and dark energy, are the of the
same nature, but their act at different scales. This way of
understand the SNIa observations, implies that cosmology have a new
window to explore the universe considering matter creation, encoded
in the function $dN/dV$. Although I have discussed some examples for
this function, neither has been obtained from a fundamental
theoretical basis.

\section*{Acknowledgments}

The author want to thank S. del Campo and R. Herrera for useful
discussions.


\end{document}